\documentstyle[11pt]{article}

\newcommand{\tfrac}[2]{{\textstyle{#1\over#2}}}
\newcommand{\dfrac}[2]{{\displaystyle{#1\over#2}}}

          \title{
                 Non-abelian bosonization from factored coset models
                 in path integrals}
          \medskip
          \author{
          A.N. Theron$^a$, F.A. Schaposnik$^b$, F.G. Scholtz$^a$ and
          H.B. Geyer$^a$\\[24pt]
          {\sl $^a$Institute of Theoretical Physics, University of
          Stellenbosch,}\\
          {\sl Stellenbosch 7600, South Africa}\\[8pt]
          {\sl $^b$Departamento de Fisica, Universidad Nacional de La
               Plata,}\\
          {\sl C.C. 67, 1900 La Plata, Argentina}
          }
\begin{document}

\maketitle

\begin{abstract}
We present a derivation of abelian and non-abelian bosonization in a
path integral setting by expressing the generating functional for
current-current correlation functions as a product of a $G/G$-coset
model, which is dynamically trivial, and a bosonic part which
contains the dynamics.  A BRST symmetry can be identified which leads
to smooth bosonization in both the abelian and non-abelian cases.
\end{abstract}

\section{Introduction}

Bosonization has proven to be one of the most important aspects
of quantum field theory in two dimensions and
after the work of Coleman \cite{coleman} and
Mandelstam \cite{mandelstam} the formalism was expanded and
refined. These results are typically summarized in what has
become known as a bosonization dictionary which captures the
one-to-one relations between elements of the fermionic and bosonic
descriptions and allows one to translate a fermionic theory into a
bosonic theory.  The dictionary equates for example the free
massless Dirac field with a free massless bosonic field and it gives
rules for how some fermionic operators can be replaced by bosonic
ones. At the same time it provides a powerful tool to analyze
fermionic models in two dimensions.  A non-abelian extension of this
approach has been given by Witten \cite{witten}.

The versatility of path integrals clearly indicate the
desirability to develop these techniques also in a
path integral language.  A path integral version of the bosonization
technique for interacting models (like the Thirring and
Schwinger models) had in fact been developed some time
ago \cite{gss}, but it was only recently \cite{dns} that (in the
abelian case) the equivalence between free bosons and free fermions
was established with path integral techniques.

The central idea in ref.\  \cite{dns} was to introduce bosonic degrees
of freedom into the path integral in such a way that a gauge symmetry
appears, so that the fermionic degrees of freedom can then
be ``frozen" out by fixing a gauge.  Although conceptually appealing,
this idea seems somehow difficult to implement in practice.

The aim of the present paper is to derive a bosonization
dictionary from path integrals in an alternative way.
Fermionic degrees of freedom
are traded for bosonic degrees of freedom by rewriting the
generating functional of the fermions in such a way that a fermionic
$G/G$ coset model \cite{rs} is factored out.  This is achieved by
first introducing a suitable identity followed by a chiral
change of variables.

This method offers a number of advantages over the method of
ref.\ \cite{dns}.  Firstly, it is technically much simpler, mainly
because no gauge fixing is necessary and therefore the complications
due to gauge fixing in \cite{dns} are avoided.  More importantly, our
method generalizes readily to the non-abelian case.  Indeed, for
fermions in a representation of a non-abelian group $G$, we are able
to show the equivalence between a theory of free fermions and a $G/G$
coset model times a Wess-Zumino-Witten (WZW) model \cite{witten}.
In this way we rederive in a rather straightforward way Witten's
non-abelian bosonization \cite{witten} for free fermions in the path
integral setting.  Furthermore our method enables us to derive a
complete bosonization dictionary in path integral language.

Concerning the $G/G$ coset model \cite{gko} which factors out in our
approach, we note that it corresponds to a topological model, so
that the associated energy momentum tensor is zero -- no
dynamical excitations exist in this sector.  The original fermionic
model is thus factored into a dynamically trivial model and
a bosonic model which carries the physical degrees of freedom of the
original fermions.

It should be noted that we concentrate on correlation functions
of currents and other fermionic bilinear products.  For
simplicity we do not consider energy-momentum correlation functions,
but our method can be extended to study such correlation functions
in a straigthforward way by starting with free fermions in a
background metric.  Here  we employ a flat Euclidian space-time
throughout.

Although our method is logically independent from the method in
ref.\ \cite{dns}, there are interesting parallels.  In particular,
although we do not introduce a bosonization gauge symmetry from the
start as in ref.\ \cite{dns}, we discover the existence of a
BRST-symmetry which appears naturally in our work.  This BRST symmetry
is especially important in the non-abelian case where it guides us in
handling a determinant.

The outline of the paper is as follows.  In Sect.\ 2 we discuss our
boson\-ization technique in detail in the abelian case.  In
Sect.\
3 we discuss the BRST symmetry and some parallels with the work in
ref.\  \cite{dns}.  The general formalism for the non-abelain case is
presented in Sect.\ 4 and in Sect.\ 5 we summarize and draw some
conclusions.

\section{Abelian bosonization}

For simplicity we first illustrate the method in the abelian
case.  We consider the problem of free Dirac fermions in two
dimensional Euclidian space for which the generating functional
for current-current correlation functions is
\begin{equation}
\label{2.1}
  Z[A_{\mu} ]
  =   \int \! D \bar{\psi} D \psi e^{-\int \! d^2 x \,[\bar{\psi}
 i \gamma_{\mu} \partial_{\mu} \psi - j_{\mu} A_{\mu}]  }   \; .
\end{equation}
In holomorphic and anti-holomorphic coordinates this reads
\begin{equation}
\label{2.2}
   Z[A_{+},A_{-} ] =   \int \! D \bar{\psi} D \psi e^{-\int \! d^2 x\,
[   \psi_{-}^{\dag}
  i \partial_{+} \psi_{-} + \psi_{+}^{\dag}
 i \partial_{-} \psi_{+}
   -A_{-}j_{+} -  A_{+}j_{-} ] } \; ,
\end{equation}
but it is general enough to consider
\begin{equation}
\label{2.3}
   Z[A_{+}] =   \int \! D \bar{\psi} D \psi e^{-\int \! d^2 x\,[
   \psi_{-}^{\dag}
  i \partial_{+} \psi_{-} + \psi_{+}^{\dag}
  i  \partial_{-} \psi_{+}
    -  A_{+}j_{-}]  } \quad.
\end{equation}
This becomes transparent once we make a  change of variables in
Eq.\  (\ref{2.2}), namely
 \begin{eqnarray}
 \label{2.4}
      \psi^{\prime}  & = &  e^{i \eta (x) } \psi \nonumber \\
      \bar {\psi}^{\prime} & = & \bar{\psi}   e^{- i \eta (x) }
       \quad,
   \end{eqnarray}
where $\eta(x)$ is chosen so that $\partial_{-} \eta (x) =- A_{-}$
and where we introduce $A^{\prime}_{+} = A_{+} + \partial_{+} \eta$.
The fermionic measure is invariant under this transformation as
it is a gauge transformation and Eq.\ (\ref{2.2}) becomes
 \begin{equation}
\label{2.5}
    Z[A_{+}^{\prime} ] =   \int \! D \bar{\psi}^{\prime} D \psi^{\prime}
e^{-\int \!
    d^2 x\,[ \psi_{-}^{ \prime \dag }
 i  \partial_{+} \psi_{-}^{\prime} + \psi_{+}^{ \prime \dag  }
 i  \partial_{-} \psi_{+}^{\prime}
    -  A_{+}^{\prime}    j_{-}^{\prime}]  } \quad.
\end{equation}
Dropping primes for notational simplicity, this is just
expression (\ref{2.3}) as asserted.  For the sake of technical
convenience we use expression (\ref{2.3}) rather than (\ref{2.2}) and
at the end of the section re-introduce the original sources and
discuss how to bosonize without making transformation (\ref{2.4}).

The key idea to our approach is to rewrite the generating
functional (\ref{2.5}) in the form

\begin{equation}
\label{2.6a}
  \int \! D \bar{\psi} D  \psi D B_{+} e^{-S_{\rm{cf}} }\times \int \! D \phi
  \, e^{-S_{\rm {B}}[\phi ,A_{+}]}
  \quad,
\end{equation}
where    $S_{\rm{cf}}$ is the action of a constrained fermionic
$U(1)/U(1)$
coset model which in light cone gauge reads (see Refs.
\cite{rs,pol}) %
\begin{equation}
\label{2.6}
   S_{\rm{cf}} = \int \! d^2 x\,[ \psi_{-}^{\dag}
    i \partial_{+} \psi_{-} + \psi_{+}^{\dag}
     i \partial_{-} \psi_{+}
      -  B_{+}j_{-} + \rm{ghosts}]    \quad,
\end{equation}
with $B_{+}$ in the role of a Lagrange multiplier that
enforces the coset condition on physical states \cite{gko,pol}.
$S_{\rm{B}}$ is a bosonic action
to be determined and the source term $A_{+}$ is
coupled only to the bosons of this model.

To achieve this we have to introduce a bosonic field into the generating
functional (\ref{2.5}) which will play the role of the field $B_{+}$ in
(\ref{2.6}). With this in mind, we insert the identity
\begin{equation}
\label{2.8}
   1=\int \! D B_{+}  e^{ \int \! d^2 x\, B_{+} j_{-} }  \delta( B_{+})
\end{equation}
into the generating functional (\ref{2.5}) which now becomes
\begin{equation}
\label{2.9}
   Z[A_{+}] =   \int \! D \bar{\psi} D \psi D B_{+}
    \delta(B_{+})
     e^{-\int \!    d^2 x\,[ L_{\rm{F}}
    -  A_{+}    j_{-}  - B_{+}    j_{-}]  } \quad.
  \end{equation}
where we have dropped the primes and introduced
\begin{equation}
L_{\rm{F}} =  \psi_{-}^{\dag }  i \partial_{+} \psi_{-} + \psi_{+}^{\dag  }
  i \partial_{-} \psi_{+}  \quad.
  \end{equation}
We  represent  the delta   functional as
\begin{equation}
   \delta(B_{+}) =\int \! D \lambda_{-}  e^{\int \! d^2 x\, B_{+} \lambda_{-}
} \end{equation}
and shift the field $B_{+}$ to
$B_{+}^{\prime}=B_{+}+A_{+}$,
so that Eq.\  (\ref{2.9})  becomes
\begin{equation}
\label{2.10}
   Z[A_{+}] =  \int \! D \bar{\psi} D \psi D
                B_{+}^{\prime} D \lambda_{-} e^{-\int \!
                d^2 x\,  [ L_{\rm{F}} - B_{+}^{\prime}    j_{-}
                - B_{+}^{\prime} \lambda_{-}  + A_{+} \lambda_{-}]  }
                \quad.
\end{equation}

Comparing $Z[A_{+}]$ above with Eqs.\ (\ref{2.6a}) and
(\ref{2.6}), we see that the first two terms of the action are
in fact those of a constrained fermionic model.  In order to
cast Eq.\ (\ref{2.10}) completely in the form of Eq.\
(\ref{2.6}), the third term in Eq.\ (\ref{2.10}) will have to
vanish. This can be achieved by a chiral change of
variables in Eq.\ (\ref{2.10}), since a term linear in
$B_{+}^{\prime}$ is then generated as a result of the
chiral anomaly which originates from the non-trivial way the fermionic
measure transforms under a chiral rotation \cite{fuj}.
To see this, consider first an infinitesimal transformation
\begin{equation}
   \label{2.11}
   \begin{array}{rclcrcl}
     \psi^{\prime}_{-}  & = &  e^{i \theta (x) } \psi_{-} &\quad
     \quad&
     \psi^{\prime}_{+}  & = &   \psi_{+}
     \nonumber \\
     \psi^{\prime \dag }_{-} & = & \psi^{\dag}_{-}   e^{- i \theta (x) }
     & \quad\quad & \psi^{\prime \dag}_{+} & = &  \psi^{\dag}_{+}
    \quad,
  \end{array}
\end{equation}
which yields the Fujikawa Jacobian \cite{fuj}
\begin{equation}
   \label{2.11a}
J_{\rm{F}}= e^{\frac{i}{2 \pi} \int\!d^2 x \,  \theta \tilde{F}}
\end{equation}
where
\begin{equation}
\label{2.12}
\begin{array}{ll}
 \tilde{F}& = \epsilon_{\mu \nu} \partial_{\mu} B_{\nu}^{\prime}\nonumber \\
          &  =- 2i \partial_{-}B_{+}^{\prime} \qquad.
   \end{array}
\end{equation}
$B_{\mu}^{\prime}$ is the field which couples to the current
$j_{\mu}$ and appears in the covariant
derivative used to regularize the Fujikawa Jacobian.

Extension to a finite chiral rotation yields the Jacobian
\begin{equation}
\label{2.12a}
J_{\rm{F}} =  \exp\bigl(\frac{1}{ \pi}  \int\!d^2x\,    (
\tfrac{1}{2} \theta
          \partial_{-} \partial_{+} \theta  + \theta  \partial_{-}
B_{+}^{\prime})\bigr)\; .
\end{equation}
This result is more complicated than the corresponding expression
(\ref{2.11a}) for an infinitesimal transformation, because under a
finite transformation the field $B_{\mu}^{\prime}$, which is coupled
to $j_{\mu}$ and is used as a regulator, changes.
Before exploiting expression (\ref{2.12a}) to address the unwanted
linear term in expression (\ref{2.10}), we present a short derivation
of $J_{\rm F}$. Introduce a chiral transformation
\begin{eqnarray}
   \label{2.13}
     \psi_{-}  & \rightarrow &   e^{i \alpha     \theta (x) } \psi_{-}
\nonumber \\
     \psi^{\dag}_{-} & \rightarrow & \psi^{\dag}_{-}   e^{- i \alpha
       \theta (x) }        \quad ,
\end{eqnarray}
with $\alpha$  a real number between 0 and 1, to obtain
\begin{eqnarray}
&&\psi_{-}^{\dag}
i\partial_{+} \psi_{-} + \psi_{+}^{\dag}
i\partial_{-} \psi_{+}
-B_{+}^{\prime}j_{-} \nonumber\\
& \rightarrow &
\psi_{-}^{\dag}
i\partial_{+} \psi_{-} + \psi_{+}^{\dag}
i\partial_{-} \psi_{+}
-(B_{+}^{\prime}  +\alpha \partial_{+} \theta) j_{-}\quad.
\end{eqnarray}
For the associated infinitesimal transformation
\begin{eqnarray}
     \psi_{-}  & \rightarrow &  e^{i\Delta \! \alpha     \theta (x) }
     \psi_{-} \nonumber \\
     \psi^{\dag}_{-} & \rightarrow & \psi^{\dag}_{-}   e^{- i \Delta
     \! \alpha \theta (x) }
  \end{eqnarray}
we correspondingly substitute $B_{+}^{\prime}\rightarrow
B_{+}^{\prime} +\alpha \partial_{+} \theta$ in Eq.\ (\ref{2.12}) to
obtain the contribution
\begin{equation}
    J_{\rm{F}}=\exp\Bigl(\frac{ 1}{ \pi} \int\!  d^2 x    (  \Delta\!
               \alpha \partial_{-}(B_{+}^{\prime} + \alpha
                \partial_{+} \theta ))\Bigr) \quad.
\end{equation}
Summing over $\Delta\alpha$ now gives the contribution for the finite
transformation,
\begin{eqnarray}
\label{2.15}
J_{\rm{F}} & = & \exp\bigl(\frac{ 1}{ \pi} \int \!d^2x   \int_0^1 \!
               d \alpha \,   ( \alpha \theta
              \partial_{-} \partial_{+} \theta  + \theta  \partial_{-}
               B_{+}^{\prime})\bigr) \nonumber \\
          & = & \exp\Bigl(\frac{1}{\pi}\int \!d^2x(
                \tfrac{1}{2} \theta\partial_{-} \partial_{+} \theta  + \theta
               \partial_{-} B_{+}^{\prime})\Bigr)\; ,
\end{eqnarray}
as asserted.

Returning to Eq.(\ref{2.10}) we see that a finite
transformation
(\ref{2.11}) for $\theta$ in Eq.(\ref{2.10}) yields
\begin{eqnarray}
\label{2.16}
     Z[A_{+}] & = &  \int \! D \bar{\psi}^{\prime} D \psi^{\prime} D
               B_{+}^{\prime}                D \lambda_{-}      e^{-
               S^{\prime}}    \nonumber \\
    S^{\prime}& = & \int \! d^2 x\,[ \psi_{-}^{\dag \prime  }
         i \partial_{+} \psi_{-}^{\prime} + \psi_{+}^{\dag \prime   }
         i \partial_{-} \psi_{+}^{\prime}
        -( B_{+}^{\prime} +\partial_{+}\theta)   j_{-}^{\prime}
         \nonumber \\
     & &
     +\frac{1}{2 \pi} \partial_{-}\theta \partial_{+} \theta   + \frac{1}{\pi}
     B_{+}^{\prime}  \partial_{-} \theta
   - B_{+}^{\prime} \lambda_{-}  + A_{+} \lambda_{-} ]\quad.
\end{eqnarray}

We are now almost in a position to show the equivalence of
expressions (\ref{2.10}) and (\ref{2.6a}).
By making the change of variables
\begin{equation}
\label{change}
\lambda_{-} =     - \frac{1}{\pi} \partial_{-} \theta ,
\end{equation}
the second and third last terms cancel  and we obtain
\begin{equation}
\label{2.17}
   Z[A_{+} ] =   \int \! D \bar{\psi} D \psi D
                B_{+}^{\prime} D \theta\, {\rm det}(\frac{1}{\pi}
                \partial_{-})  e^{-\int \!
        d^2 x\,L}
\end{equation}
with
\begin{equation}
L= L_{\rm F}
        -( B_{+}^{\prime} +\partial_{+} \theta)   j_{-}
       +\frac{1}{2 \pi} \partial_{-} \theta \partial_{+} \theta    +
       \frac{1}{\pi}
       A_{+}  \partial_{-}\theta\; ,
\end{equation}
while the determinant   originating from the change of
variables (\ref{change})  can be represented by ghosts in the
form
$  \int \! D     \bar{c} D c  e^{-\int \! d^2 x \, \bar{c}
\partial_{-} c}$.

Finally, shifting $B_{+}^{\prime} \rightarrow B_{+}^{\prime} -
\partial_{+} \theta$, we see that we have succeeded in our
original aim: the free fermion generating functional has been written
as the product of partition functions for a coset model
and a theory of free bosons.

A point to note is that the coset model obtained above is in the light
cone gauge.  Normally the $U(1)/U(1)$ coset model is formulated in
terms of fermions as \cite{rs,pol}
\begin{equation}
\label{2.21}
   L_{U(1)/U(1)}  =   \psi_{-}^{\dag}
   \partial_{+} \psi_{-} + \psi_{+}^{\dag}
  \partial_{-} \psi_{+}
   +B_{-}j_{+} +  B_{+}j_{-}   \quad.
\end{equation}
This model has a $U(1)$ gauge invariance which needs to be gauge
fixed. If we choose the light cone gauge we have
\begin{equation}
\label{2.22}
\begin{array}{ll}
     Z_{U(1)/U(1)}  & =   \int \! D \bar{\psi} D \psi D B_{+}
     D B_{-}  D \bar{c} D c  \delta(B_{-}) \Delta_{\rm{FP}}  e^{-\int \!
d^2 x
           L_{U(1)/U(1)} }    \nonumber  \\
        & =   \int \! D \bar{\psi} D \psi D  B_{+}      e^{-\int \!
  d^2 x \, [  L_{U(1)/U(1)}
+ B_{+}    j_{-}    +  \bar{c} \partial_{-} c    ]  } \quad,
  \end{array}
\end{equation}
where $\Delta_{\rm{FP}}$ is the Fadeev-Popov determinant.
In the second line we have represented this determinant with ghosts.
We note that the fermionic parts of expressions (\ref{2.17}) and
(\ref{2.22}) are the same. The ghosts introduced to lift the
Jacobian in
expression (\ref{change})  may be identified with the Fadeev-Popov
ghosts in expression  (\ref{2.22}).

We now discuss how the product representation (\ref{2.17}) of the
fermionic generating functional can be used to extract the
bosonization dictionary.  To calculate current-current correlation
functions, we only need to consider the bosonic part of Eq.\
(\ref{2.17}), because the source is decoupled from the fermions in
 \begin{equation}
   Z[A_{+}^{\prime} ] = N  \int \! D \theta
        e^{-\int \!        d^2 x \,[
       +\frac{1}{2 \pi} \partial_{-} \theta \partial_{+} \theta    +
\frac{1}{\pi}
      A_{+}  \partial_{-}\theta ]  }  \quad.
  \end{equation}

We have included in the normalization $N$ the fermionic coset factor
which, as stressed above, is irrelevant for current-current
correlation functions.  In order to rewrite our result in terms of the
original sources of the generating functional (\ref{2.2}), we
substitute $A_{+}$ in the above equation by $ A_{+} + \partial_{+}
\eta$ where $\partial_{-} \eta (x) =- A_{-}$ and perform a partial
integration.  This gives
\begin{equation}
\label{2.23}
   Z[A_{+},A_{-} ] = N  \int \! D \theta
        e^{-\int \!     d^2 x \, [
       \frac{1}{2 \pi} \partial_{-} \theta \partial_{+} \theta    +
       \frac{1}{\pi}
       A_{\mu} \epsilon_{\mu \nu}  \partial_{\nu}\theta]}\; ,
\end{equation}
so that we can associate the fermionic current with $\epsilon_{\mu
\nu} \partial_{\nu}\theta$.  In order to write the bosonization rule
for the fermionic current with the usual normalization, we rescale the
bosonic field according to $\theta/\sqrt{\pi} \to \theta$, so that we
find from Eq.\ (\ref{2.23}) the well known bosonization rule
\begin{equation}
\label{bosonization}
j_{\mu} \to (1/\sqrt{\pi}) \epsilon_{\mu \nu} \partial_{\nu} \theta   \quad.
\end{equation}

We have thus been able to rewrite the partition function for
free fermions in the presence of an external source as the partition
function for free bosons (times a coset model factor which plays no
physical role). By just considering the coupling of the bosonic
field to the source we could then derive the bosonization rule
(\ref{bosonization}).  Although our procedure may seem complicated for
the simple abelian case, we shall see below that it is easily extended
to the non-abelian case for the derivation of more involved
non-abelian bosonization rules.

Concerning the bosonization rules for $\psi$, we just add source terms
of the form $\psi^{\dag}_{-}\psi_{+}$ and $\psi^{\dag}_{+} \psi_{-}$
to the original fermion generating functional, extending it to
\begin{equation}
\label{2.24}
     Z[A_{+},\rho,\tilde{\rho}] = \int \! D\bar{ \psi}  D \psi e^{-\int
     \! d^2 x\,
     [ L_{\rm{F}} - A_{+} j_{-} + \rho    \psi^{\dag}_{-}\psi_{+} +
     \tilde{\rho} \psi^{\dag}_{+} \psi_{-}]  } \quad.
\end{equation}
We then go through the same steps as above to obtain Eq.\
(\ref{2.17})
with additional source terms
\begin{equation}
   Z[A_{+},\rho,\tilde{\rho} ] =   \int \! D \bar{\psi} D \psi D
         B_{+}^{\prime} D \theta
        e^{-\int \!
        d^2 x\, [ L_{\rm{CF}}  + L_{\rm{B}}
        +   \rho e^{-i \theta}
\psi^{\dag}_{-}\psi_{+} + \tilde{\rho}  e^{ i \theta}   \psi^{\dag}_{+}
\psi_{-}]  }   \quad.
  \end{equation}
If we now differentiate with respect to the sources and set them equal
to zero,
we again obtain a factorization into a coset model and a free bosonic
model. As an example we calculate
\begin{equation}
\label{2.25}
\begin{array}{ll}
&\langle\; \psi^{\dag}_{-}\psi_{+}(x_1)  \psi^{\dag}_{+}
\psi_{-}(x_2)\;\rangle_{\rm{free\,\, fermion}}\nonumber \\
& \quad \quad =
\langle\; \psi^{\dag}_{-}\psi_{+}(x_1)  \psi^{\dag}_{+}
\psi_{-}(x_2)\;\rangle_{\rm{coset}}
\langle\;e^{- i \theta(x_1)}e^{ i \theta(x_2)}\;\rangle_{\rm{free\,\,
boson}}\,\,. \end{array}
\end{equation}
The correlator $\langle\;\;\rangle _{\rm{coset}}$ is in fact a
constant \cite{pol} because, as already stated, the coset model
has no dynamical degrees of freedom.  Note that we have to restrict
our discussion here to gauge invariant fermionic quantities, as the
correlators of quantities like $\psi$ are not constant in the
coset model, but have a value which depends on how the gauge in
the coset model has been fixed.

{}From Eq.\ (\ref{2.25}) one can infer the bosonization rule
\begin{equation}
  \psi^{\dag}_{+}\psi_{-}  \to  e^{i \theta}  \; .
\label{bos2}
\end{equation}
At the canonical level $\psi_{+}$ is only a function of $x_{+}$
and $  \psi^{\dag}_{-}$ of $x_{-}$, which means that we may
associate $\psi_{+}$ with the holomorphic part of  $e^{i \theta}$.

The fact that we have obtained, as a matter of convenience, a
coset model in a particular gauge, may raise questions about the
generality of the result.  It is in fact possible to modify the
procedure to obtain a coset model where the gauge has not been fixed.
This is achieved by noting that in
\begin{equation}
\label{2.26}
  Z[A_{\mu} ] =   \int \! D \bar{\psi} D \psi e^{-\int \! d^2 x \,[\bar{\psi}
  \gamma_{\mu}i \partial_{\mu} \psi - j_{\mu} A_{\mu} ] }   \quad.
\end{equation}
we can make the following change of variables (which corresponds to a gauge
transformation)
\begin{equation}
  \begin{array}{ll}
     \psi  & \rightarrow   e^{i \eta (x) } \psi \nonumber \\
     \bar {\psi} & \rightarrow  \bar{\psi}   e^{- i \eta (x) }
      \quad.
   \end{array}
   \end{equation}
to obtain
  \begin{equation}
  Z[A_{\mu} ] =   \int \! D \bar{\psi} D \psi e^{-\int \! d^2 x\,[ \bar{\psi}
 i \gamma_{\mu} \partial_{\mu} \psi - j_{\mu} A_{\mu}  -\partial_{\mu} \eta
j_{\mu}] } \quad.
 \end{equation}
Since this generating functional is in fact $\eta$-independent, we
find after integrating over $\eta$ (and absorbing an infinite
normalization constant)
\begin{equation}
  \label{2.27}
   Z[A_{\mu} ] =   \int \! D \bar{\psi} D \psi \delta(\partial_{\mu}j_{\mu})
e^{-\int \! d^2 x\,[ \bar{\psi}
   i \gamma_{\mu} \partial_{\mu} \psi - j_{\mu} A_{\mu} ] }   \quad.
\end{equation}
We now substitute the delta functional
\begin{equation}
  \label{2.28}
 \delta(\partial_{\mu}j_{\mu}) =
 \int \! D B_{\mu} D \theta e^{\int \! d^2x \,[ B_{\mu}
    j_{\mu} + \frac{1}{\pi}B_{\mu}\epsilon_{\mu \nu} \partial_{\nu}
\theta] } \quad.
 \end{equation}
in Eq.\ (\ref{2.27}) to obtain
\begin{equation}
  \label{2.29}
   Z[A_{\mu} ] =   \int \! D \bar{\psi} D \psi D
B_{\mu} D \theta e^{-\int \! d^2 x\,[ \bar{\psi}
  i  \gamma_{\mu} \partial_{\mu} \psi - j_{\mu}( A_{\mu} +B_{\mu}) -
\frac{1}{\pi}B_{\mu}\epsilon_{\mu \nu} \partial_{\nu}\theta ]    }
 \quad.
 \end{equation}
(The identity (\ref{2.28}) can be obtained by first integrating
over the field $\theta$ and then over $B_{\mu}$.)

We can now obtain a constrained fermionic model by introducing
the change of variables
  \begin{eqnarray}
     \psi  & \rightarrow &   e^{i \gamma_5   \theta^{\prime} (x) }
            \psi \nonumber \\
     \bar {\psi} & \rightarrow & \bar{\psi}   e^{- i  \gamma_5
           \theta^{\prime}(x) } \nonumber \\
        \theta(x) & \rightarrow & \theta^{\prime}(x)\; ,
  \end{eqnarray}
The last term in Eq.\ (\ref{2.29}) then cancels  the linear term
in $B$ of the Fujikawa determinant so that we  finally  have
\begin{equation}
  \label{2.30}
   Z[A_{\mu} ] =   \int \! D \bar{\psi} D \psi D
B_{\mu} D \theta\, e^{-\int \! d^2 x L}
\end{equation}
with
\begin{equation}
L= \bar{\psi}
   i \gamma_{\mu} \partial_{\mu} \psi -j_{\mu}( A_{\mu} + B_{\mu}
     +\epsilon_{\mu \nu} \partial_{\nu}\theta)
     +\frac{1}{\pi}A_{\mu}\epsilon_{\mu \nu} \partial_{\nu}\theta
     +\frac{1}{2  \pi} \partial_{\nu}\theta \partial_{\nu}\theta
     \; ,
\end{equation}
where we have used $\gamma_{\mu}\gamma_5 =  \epsilon_{\mu \nu}
\gamma_{\nu}$ and partial integrations.

Eq.\ (\ref{2.30}) is the desired result.  Shifting the field $B_{\mu}
\rightarrow A_{\mu} + B_{\mu} +\epsilon_{\mu \nu}
\partial_{\nu}\theta$ we obtain a constrained fermionic model which is
not gauge fixed together with a free bosonic theory.  The
procedure is  simpler, however, if we work only with right-handed
fermions at intermediate stages, especially in the non-abelian case,
as becomes apparent in Sect.\ 4.

\section{BRST symmetry and smooth bosonization}

In this section we briefly discuss the relation between our procedure
and the smooth bosonization approach of ref.\ \cite{dns},
where a gauge symmetry, referred to as ``bosonization
gauge symmetry", was introduced by making a $\gamma_5$
transformation and integrating over the corresponding angle.  This
procedure indicates that bosonization can be viewed as arising
from the choice of a particular gauge.  Although no gauge symmetry is
introduced in our approach, a similar view of bosonization can be
taken by noting that a BRST symmetry arises naturally within our
scheme.

Consider again Eq.\ (\ref{2.10})
\begin{equation}
\label{3.1}
   Z[A_{+}] =   \int \! D \bar{\psi} D \psi D B_{+}   D \lambda_{-}
     e^{-\int \! d^2 x\,[ L_{\rm{F}} -  A_{+}    j_{-}  - B_{+} j_{-}
   - B_{+} \lambda_{-}] } \quad.
\end{equation}
Making the change of variables
\begin{equation}
   \begin{array}{rclcrcl}
     \psi^{\prime}_{-}  & = &  e^{i\theta (x)}\psi_{-} &\quad \quad&
     \psi^{\prime}_{+}  & = &   \psi_{+}
     \nonumber \\
     \psi^{\prime \dag }_{-} & = & \psi^{\dag}_{-}   e^{- i \theta (x) }
     &\quad\quad &\psi^{\prime \dag}_{+} & = & \psi^{\dag}_{+}
     \nonumber \\
     \lambda_{-} &=& \lambda_{-}(\theta) &&&&\quad,
  \end{array}
\end{equation}
we obtain
\begin{eqnarray}
\label{3.2}
     Z[A_{+} ] & = &  \int \! D \bar{\psi} D \psi D
               B_{+}  D \theta   D \bar{c} D c\, e^{-S}
               \nonumber \\
      S  &  = & \int \!  d^2 x\,[  L_{\rm{F}}
            -( B_{+} + A_{+}      + \partial_{+}\theta)   j_{-}
            +\frac{1}{2 \pi} \partial_{-}\theta \partial_{+} \theta
            \nonumber \\
   &  &  +  \frac{1}{\pi} (B_{+} + A_{+})  \partial_{-} \theta
         + B_{+} \lambda_{-}(\theta)  + \bar{c} \frac{\delta
         \lambda_{-}(\theta)}{\delta \theta} c]
\end{eqnarray}
where we have lifted the Jacobian induced by
$\lambda_{-} \rightarrow \lambda_{-} (\theta)$ with ghosts.
Apart from the last two terms the above generating functional has a gauge
invariance under
 \begin{equation}
    \label{3.3}
    \begin{array}{rclcrcl}
      \psi_{-}  & \rightarrow&   e^{i \alpha(x)  } \psi_{-}
      &\quad\quad&    \psi_{+}   &\rightarrow&    \psi_{+}
      \nonumber \\
      \psi^{\dag}_{-} & \rightarrow &\psi^{\dag}_{-}   e^{- i
      \alpha(x) } &\quad\quad&
      \psi^{\dag}_{+}  &\rightarrow & \psi^{\dag}_{+}
        \nonumber \\
       \theta(x) & \rightarrow &\theta(x) - \alpha (x)  &&&& \quad  .
   \end{array}
   \end{equation}
This can easily be verified explicitly:
 the part of the Lagrangian that contains fermions is exactly invariant under
the transformation while the variation of the bosonic part cancels the
contribution coming from the fermionic measure. Only the last
two terms break the symmetry explicitly.
The existence of this symmetry
stems from
the chiral  transformation we made and the fact that we also
integrate over the parameter of the transformation.

Although the total generating functional is not invariant under the
transformation Eq.\ (\ref{3.3}),    a  global   BRST
symmetry  exists which leaves the generating functional
invariant.
This follows once we note that the last two terms may be written
as a BRST exact form
  \begin{equation}
  \label{3.4}
\delta(\bar{c} \lambda_{-} ) =  B_{+} \lambda_{-} +   \bar{ c}\frac{\delta
\lambda_{-}}{\delta \theta} c
\end{equation}
where (with a global Grassmann number omitted for simplicity)
\begin{equation}
\begin{array}{rclcrcl}
    \delta \bar{c} & = & B_{+} &\quad\quad & \delta \psi_{-} &=& i c
    \psi_{-} \nonumber \\
    \delta c & = & 0 &\quad\quad& \delta \psi_{-}^{\dag} &=& - i
    \psi^{\dag}_{-}c \nonumber \\
    \delta \theta &=& - c  &\quad\quad& \delta \psi_{+}  &=& 0
    \nonumber \\
    \delta B_{+}&=&0  &\quad\quad&  \delta\psi_{+}^{\dag} &=& 0
    \nonumber \\
    \delta A_{+} & = & 0 &&&& \quad .
\end{array}
\end{equation}
This BRST symmetry is nilpotent and therefore the variation of Eq.\
(\ref{3.4}) is zero.  The remainder of the partition function is
invariant because this BRST symmetry is only a special case of the
symmetry (\ref{3.3}).  The terms in expression (\ref{3.4}) can
therefore be viewed as a type of gauge fixing condition and a
Fadeev-Popov determinant, respectively.  Note however that no notion
of anomalous gauge fixing as discussed in \cite{dns} is required in
our approach.

If we now choose $\lambda_{-} = - \frac{1}{\pi}
\partial_{-} \theta$
in Eq.\ (\ref{3.2}), we obtain the bosonization results of the
previous
section. We have the freedom, however, of choosing a still more
general
$\lambda_{-}(\theta)$. This will correspond to the smooth bosonization of
\cite{dns}.
As an  example, instead of the identity (\ref{2.8}) we can use
 \begin{equation}
   1=\int \! D B_{+}  e^{-\int \! d^2 x \, \alpha  B_{+} j_{-} }  \delta(
B_{+})
   \quad.
\end{equation}
where $\alpha$ is a parameter between 0 and 1. Eq.(\ref{3.2}) becomes
\begin{equation}
     Z[A_{+} ] =   \int \! D \bar{\psi} D \psi D
               B_{+}  D \theta   D \bar{c} D c \,  e^{-  S}
 \end{equation}
with
\begin{eqnarray}
S &=&    \int \!       d^2x\,[ L_{\rm{F}}
      -(\alpha B_{+} + A_{+}      + \partial_{+}\theta)   j_{-}
     +\frac{1}{2 \pi} \partial_{-}\theta \partial_{+} \theta
     \nonumber\\
     &&+ \frac{1}{\pi} (\alpha B_{+} + A_{+})  \partial_{-} \theta
      + B_{+} \lambda_{-}  + \bar{ c}\frac{\delta \lambda_{-}}{\delta
      \theta} c  ]
\end{eqnarray}
The choice $\lambda_{-} = - \frac{1}{\pi} \partial_{-} \theta$
now leads for $\alpha =1$ to a constrained fermion model and a free
boson model, and in the case $\alpha=0$ to a free fermion model and a
trivial boson theory.  For values of $\alpha$ between 0 and 1 one
finds non-trivial models of fermions and bosons coupled together.

\section{Non-abelian bosonization}

In this section we generalize our approach to bosonization to
the non-abelian case.  One of the advantages of our scheme is in
fact the simplicity with which abelian bosonization can be extended
to the case of fermions in some representation of a non-abelian group
$G$.  We start with a theory of free Dirac fermions which are in the
fundamental representation of $U(N)$. The generating functional
is
\begin{equation}
\label{4.1}
   Z[A_{+},A_{-} ] =   \int \! D \bar{\psi} D \psi e^{-\int \! d^2 x\,[
   \psi_{-}^{\dag}  i \partial_{+} \psi_{-} +
 \psi_{+}^{\dag}  i \partial_{-} \psi_{+}
   -\psi_{-}^{\dag} A_{+}  \psi_{-} - \psi_{+}^{\dag} A_{-} \psi_{+}  ]  }
\end{equation}
where $\psi$ is in the fundamental representation and
$A= A^{a}T^{a}$.

As in the abelian case we find it convenient to set one of the source
terms equal to zero at intermediate stages of the calculation.  This
can be accomplished by the following gauge transformation which we
view as a change of variables
\begin{equation}
   \label{4.3}
   \begin{array}{rclcrcl}
     \psi^{\prime}_{-}  & = &  h \psi_{-} &\quad \quad&
    \psi^{\prime}_{+}  & = &  h \psi_{+}
\nonumber \\
     \psi^{\dag  \prime}_{-} & = & \psi^{\dag}_{-}  h^{-1}
     &\quad\quad& \psi^{\dag  \prime}_{+} & = &  \psi^{\dag}_{+}
     h^{-1}\; .
  \end{array}
  \end{equation}
We choose $h$ so that
\begin{equation}
   \label{4.4}
   h^{-1}A_{-} h -i h^{-1}\partial_{-}h=0
\end{equation}
and introduce $A^{\prime }_{+}
=h^{-1}A_{+} h -i h^{-1}\partial_{+}h$.
The fermionic measure is invariant under this type of transformation.
In the following we again drop the primes for notational simplicity.

To rewrite Eq.\  (\ref{4.1}) as a functional which factorizes
into a constrained
fermionic model and a bosonic part, requires once again a field
$B_{+}$ which will become the Lagrange multiplier of the constrained fermions.
The identity (\ref{2.8}) generalizes in an obvious way to
\begin{eqnarray}
   \label{4.5}
       1 &=& \int \! D B_{+} \delta( B_{+})    e^{\int \! d^2 x \,
       \psi^{\dag}_{-} B_{+} \psi _{-} }
        \nonumber \\
      &=& \int \! D B_{+} D \lambda_{-}   e^{\int \! d^2 x \,[
        \psi^{\dag}_{-} B_{+} \psi _{-}  +Tr(B_{+} \lambda_{-})] }\; ,
\end{eqnarray}
where $B_{+}=B_{+}^{a}T^{a}$ and $\lambda_{-}=\lambda_{-}^{a}T^{a}$.

This identity is introduced in Eq.\ (\ref{4.1}) and to constrain
the fermions we perform the chiral change of variables
 \begin{equation}
   \label{4.6}
   \begin{array}{rclcrcl}
     \psi^{\prime}_{-}  & = & g \psi_{-} &\quad \quad&
     \psi^{\prime}_{+}  & = &   \psi_{+}
     \nonumber \\
     \psi^{\dag  \prime}_{-} & = & \psi^{\dag}_{-} g^{-1}
     &\quad\quad& \psi^{\dag  \prime}_{+} & = & \psi^{\dag}_{+}
     \nonumber \\
    \lambda & =& \lambda(g)  &&&& \; .
  \end{array}
  \end{equation}
The fermionic measure is not invariant under the transformation, since under an
infinitesimal transformation
\begin{eqnarray}
   \label{4.7}
     \psi_{-}  & \rightarrow &   e^{i   \theta (x) } \psi_{-}
     \nonumber \\
     \psi^{\dag}_{-} & \rightarrow & \psi^{\dag}_{-}   e^{- i
     \theta (x) }\; ,
\end{eqnarray}
where  $\theta= \theta^{a}T^{a}$, one already infers the
non-trivial Jacobian $J_{\rm{F}}= \frac{i}{2 \pi} \theta
\tilde{F}$, with
\begin{eqnarray}
 \tilde{F}& = & \epsilon_{\mu \nu} F_{\mu \nu } \nonumber \\
  F_{\mu \nu} & = & \partial_{\mu} A_{\nu}- \partial_{\nu} A_{\nu}
  +i [A_{\mu},A_{\nu}] \; .
\end{eqnarray}
Here $A_{\mu}$ is the field appearing in in the covariant
derivative
used as a regulator when the Jacobian is calculated.
In the present case $ J_{\rm{F}} = \frac{1}{\pi}\theta
\partial_{-}(B_{+} +A_{+} )$.

The corresponding determinant for a finite transformation may be
 calculated as in the abelian case.  Here we exploit the
result \cite{wieg,gss}
\begin{equation}
\label{4.9a}
{\rm det}(i \partial_{+} -A_{+} )
=e^{S[g]} \quad,
\end{equation}
where $g$ is defined by  $A_{+}= -i g^{-1} \partial_{+}g$, and  $S[g]$ is
the Wess-Zumino-Witten action
\begin{eqnarray}
\label{4.10}
  S[g] &=& - \frac{1}{8\pi}\int\!d^2 x\,{\rm Tr}(g\partial_{\mu}
         g^{-1}g\partial_{\mu} g^{-1})
        \nonumber \\
       & &
       -\frac{i}{12\pi}\int_{\Gamma}\!d^3 x\, \epsilon_{\mu \nu \rho}
       {\rm Tr}(g\partial_{\mu} g^{-1}g\partial_{\nu}g^{-1}
       g\partial_{\rho} g^{-1})\quad.
\end{eqnarray}
($\Gamma$ is a three dimensional ball with two
dimesional spacetime as boundary.) This action satisfies the
Polyakov-Wiegmann identity \cite{wieg}
\begin{equation}
\label{4.11}
S[hg]= S[h] + S[g]   -  \frac{1}{4\pi}\int\!d^2x \, {\rm
       Tr}[h^{-1}(\partial_{+}  h)(\partial_{-}g) g^{-1}] \quad.
\end{equation}
The Jacobian $J$ appearing in
\begin{eqnarray}
\label{4.8}
   &&   \int \! D \bar{\psi} D \psi e^{-\int \! d^2 x\,[
   \psi_{-}^{\dag}
  i \partial_{+} \psi_{-} + \psi_{+}^{\dag}
 i \partial_{-} \psi_{+}
-  \psi_{-}^{\dag}
(A_{+} + B_{+}) \psi_{-}] }
\nonumber \\
  & = & \int \! D \bar{\psi} D \psi  J    e^{-\int \! d^2 x\,[
  \psi_{-}^{\dag}g^{-1}i  \partial_{+}(g \psi_{-}) + \psi_{+}^{\dag}
i \partial_{-} \psi_{+}
-\psi_{-}^{\dag}g^{-1}(A_{+}+ B_{+})g \psi_{-}] }
  \end{eqnarray}
can then be determined after integrating over the fermions,
namely
\begin{eqnarray}
\label{4.9}
   J &=& \frac{{\rm det}( i\partial_{+} -(A_{+}+ B_{+}))  }{{\rm det}
         (i\partial_{+}
         -(g^{-1} A_{+}g+ g^{-1}B_{+}g) +g^{-1}i\partial_{+}g )}
         \nonumber \\
     &=& \frac{e^{S[h]}}{e^{S[hg]}}\nonumber \\
     &=& e^{S[h]-S[hg]}    \nonumber \\
     &=& e^{-S[g]+\frac{i}{4\pi}\int\!d^2 x\,{\rm Tr}(( A_{+}+
         B_{+})(\partial_{-} g)g^{-1})}
\end{eqnarray}
where $A_{+}+ B_{+}= -i h^{-1} \partial_{+} h$.

{}From the change of variables (\ref{4.6}) we therefore
obtain
\begin{equation}
\label{4.12}
     Z[A_{+} ] =   \int \! D \bar{\psi} D \psi D
               B_{+}  D g \,\, {\rm det}(\frac{\delta \lambda_{-}}{\delta g})
               \, e^{-S}
\end{equation}
with
\begin{equation}
\label{4.12a}
\begin{array}{rcll}
S & = &  S_{\rm{F}}   +S[g]
     + \int\!d^2x &\!\!\!\![  \psi_{-}^{\dag}(g^{-1}     B_{+}  g
+g^{-1} A_{+}g -i g^{-1} \partial_{+}g) \psi_{-}\nonumber\\[3pt]
&&&\!\! - \dfrac{i}{4\pi}{\rm Tr}((A_{+}+ B_{+})(\partial_{-}g)g^{-1})
   + {\rm Tr}(B_{+} \lambda_{-}(g))]
\end{array}
\end{equation}
where  ${\rm det}(\frac{\delta \lambda_{-}}{\delta g})$is the
Jacobian associated with the transformation
from $D\lambda_{-}$ to the invariant measure $Dg$, while $S_{\rm{F}}$
is the free fermion action.

As in the abelian case this generating functional is ``nearly" gauge
invariant.  Apart from the last term in the action and the determinant
${\rm det}(\frac{\delta \lambda_{-}}{\delta g})$ we have invariance
under the transformation
 \begin{equation}
    \begin{array}{rclcrcl}
  \label{gauge}
      \psi_{-}  & \rightarrow & h \psi_{-} &\quad\quad&
      \psi_{+}  & \rightarrow &   \psi_{+}
      \nonumber \\
      \psi^{\dag}_{-} & \rightarrow & \psi^{\dag}_{-}  h^{-1}
       &\quad\quad&
      \psi^{\dag}_{+}& \rightarrow & \psi^{\dag}_{+}
        \nonumber \\
       g & \rightarrow & gh^{-1} &&&& .
   \end{array}
   \end{equation}

It is straightforward to check that the part of the Lagrangian that
contains fermions is invariant, while the variation of the
bosonic part precisely cancels the anomaly arising from the
noninvariance of the fermionic measure.  The total generating
functional is, however, not gauge invariant because of the term
${\rm Tr}(B_{+} \lambda_{-}(g))$ and the above mentioned determinant,
which break the gauge symmetry explicitly.

A BRST invariance does, however, exist. As in the abelian case
the part of the generating functional which is not gauge invariant can
be written as a BRST exact form.  To see this, note that the
determinant ${\rm det}(\frac{\delta \lambda_{-}}{\delta g})$ is, like
the familiar Fadeev-Popov determinant, invariant under group
transformations
\begin{equation}
D \lambda_{-} = {\rm det}(\frac{\delta \lambda_{-}}{\delta g}) D g
=       {\rm det}(\frac{\delta \lambda_{-}}{\delta g^ {\prime }}) Dg^ {\prime }
\quad,
\end{equation}
because of the invariance of the Haar measure, and can therefore be evaluated
close to the identity.
We may parametrize $g$ by
$g=e^{i\alpha^{a} T^{a} }$ and close to the  identity  the invariant measure
is $D \alpha^{a}$.
 The determinant for the change of variables
from $\lambda_{-}$ to $\alpha$ is then ${\rm det}(\frac{\delta
\lambda_{-}}{\delta \alpha})$.
This determinant is now lifted with ghosts.
The complete generating functional is now invariant under the set
of transformations
\begin{equation}
\begin{array}{rclcrcl}
\delta \bar{c}& = & B_{+} &\quad\quad & \delta \psi_{-}  &=& i c
                  \psi_{-} \nonumber \\
\delta c & = & i c^2 &\quad\quad& \delta \psi_{-}^{\dag} &=& - i
                  \psi^{\dag}_{-}c \nonumber \\
\delta g &=& -ig c  &\quad\quad& \delta \psi_{+}  &=& 0
                  \nonumber \\
\delta g^{-1} &=& i c g^{-1} &\quad\quad& \delta\psi_{+}^{\dag} &=& 0
                  \nonumber \\
\delta B_{+}& = & 0 &\quad\quad& \delta A_{+}  &=& 0 \; .
\end{array}
\end{equation}
 In terms of the $\alpha$ this reads $\delta
\alpha =-c$ and  we have
\begin{equation}
\delta({\rm Tr} (\bar{c} \lambda_{-}))= {\rm Tr}(B_{+} \lambda_{-}  +  \bar{c}
\frac{\delta \lambda_{-}}{\delta \alpha} c)\quad,
\end{equation}
which is exactly the gauge non-invariant part.  Because the BRST
symmetry is nilpotent the BRST variation of this term is zero.  The
remainder of the generating functional is invariant because the BRST
transformations is a special case of the gauge transformation
(\ref{gauge}).  The fermionic generating functional can therefore be
written as
\begin{eqnarray}
      Z[A_{+} ] = &  \int \! D \bar{\psi} D \psi D
                B_{+}  D g   D \bar{c} D c\, e^{-S}\nonumber\\
\end{eqnarray}
with
\begin{equation}
\label{4.12b}
\begin{array}{rcll}
S &=&  S_{\rm{F}}   +S[g]
      + \int\!d^2x &\!\!\!\![  \psi_{-}^{\dag}(g^{-1} B_{+}  g
      +g^{-1} A_{+}g -i g^{-1} \partial_{+}g) \psi_{-}\nonumber\\[3pt]
      &&& \!\!- \dfrac{i}{4\pi} {\rm Tr}(( A_{+}+ B_{+})
      (\partial_{-}g)g^{-1})
      + \delta({\rm Tr}( \bar{c} \lambda_{-}))] \quad.
\end{array}
\end{equation}
For a general  $\lambda_{-} = \lambda_{-}(g)$ we have a complicated
interacting fermion and boson  model, but we can factor out a
constrained  fermionic model if we  choose
\begin{equation}
\lambda_{-}(g)=\frac{i}{4 \pi}(\partial_{-}g) g^{-1}
\end{equation}
to cancel the linear term $B_{+}(\partial_{-} g)g^{-1}$ in
Eq.\ (\ref{4.12b}).
For this choice of $\lambda_{-}$ we have
\begin{equation}
\delta({\rm Tr} (\bar{c} \lambda_{-})) = {\rm Tr}(B_{+} \lambda_{-})+
\frac{1}{4 \pi} {\rm Tr}(g^{-1} \bar{c} g \partial_{-} c)\quad. $$
Making the change of variables   $\bar{c}^{\prime}=g^{-1}\bar{c} g$, a
transformation with trivial determinant, the ghosts and bosons decouple and
Eq.\  (\ref{4.12a}) simplifies to
\begin{equation}
\label{4.13}
Z[A_{+} ] =   \int \! D \bar{\psi} D \psi D
               B_{+}  D g   D \bar{c}^{\prime} D c \, e^{ -S}
\end{equation}
with
\begin{equation}
\begin{array}{rcll}
S &=& S_{\rm{F}} +S[g]
+\int \!  d^2 x &\!\!\!\![\psi_{-}^{\dag}(g^{-1} B_{+}  g +g^{-1}
       A_{+}g -i g^{-1} \partial_{+}g)    \psi_{-}\nonumber\\[3pt]
&&&\!\! - \dfrac{i}{4\pi}{\rm Tr}(( A_{+})(\partial_{-} g)g^{-1})
   + \bar{c}^{\prime} \partial_{-}c)] \; .
\end{array}
\end{equation}
We now introduce a new Lagrange multiplier
\begin{equation}
B^{\prime}=
g^{-1}     B_{+}  g +g^{-1}  A_{+}g - ig^{-1}  \partial_{+}g\quad,
\end{equation}
a transformation for which the Jacobian is again trivial.
We then obtain the final result
\begin{equation}
\label{4.14}
      Z[A_{+} ] =   \int \! D \bar{\psi} D \psi D
               B_{+}  D g   D \bar{c}^{\prime} D c \, e^{-S}
\end{equation}
with
\begin{equation}
S = S_{\rm{F}} + S[g]+ \int \!  d^2 x [
 \psi_{-}^{\dag}  B_{+}^{\prime}\psi_{-}  + \bar{c}^{\prime}
\partial_{-} c
- \dfrac{i}{4\pi}{\rm Tr}( A_{+}(\partial_{-} g)g^{-1})]
\; .
\end{equation}

In the literature \cite{pol} one finds the Jacobian ${\rm
det}(i\partial_{+}+i g^{-1}\partial_{+}g)$ associated with the
change of variables from $DA_{+}$ to $Dg$ (where $A_{+}=
-ig^{-1}\partial_{+}g$) .  This is formally equivalent to our result,
but in the case of \cite{pol} the determinant is regularized to
respect a gauge symmetry.  In our case $\lambda$ is not a gauge
connection and there is no reason to regularize so as to respect gauge
invariance. Instead we have a global BRST symmetry
 which should be  maintained.

The interpretation of Eq.\ (\ref{4.14}) is as follows.  The fermionic
part has decoupled from the bosonic part and it is a $G/G$ coset model
which is trivial from the dynamical point of view.  All the dynamics
now resides in the bosonic sector which corresponds to a WZW
model at level one with the source of the fermionic current now
coupled to the bosonic current $(\partial_{-} g)g^{-1}$.  As in the
abelian case, we immediately infer the bosonization rule
\begin{equation}
j_{-} \to i (\partial_{-} g)g^{-1}\quad.
 \label{bos3}
\end{equation}
for the $j_{-}$ component of the fermionic current.

This model can be rewritten in terms of the original sources by making
the vector gauge transformation $hgh^{-1}$and using the invariance of
the Haar measure.  For the bosonic part of the action we find
\begin{equation}
 \label{4.16}
      Z[A_{+} ] =   \int \!                   D g
  e^{ - S[g] -           \frac{i}{4\pi}\int \!  d^2 x \, [
{\rm Tr}(A_{+}(\partial_{-} g)g^{-1})+
{\rm Tr}(A_{-}g^{-1}\partial_{+} g)  + A_{+}gA_{-}g^{-1}    +
A_{+}A_{-}] }
 \end{equation}
in terms of the original sources.

To calculate correlators of fermionic bilinears like
$\psi^{\dag}_{-}\psi_{+}$ and $\psi^{\dag}_{+} \psi_{-}$, we
simply add the corresponding source terms to Eq.\ (\ref{4.1}).
Under transformation (\ref{4.6}) these terms transform to
$\psi^{\dag}_{+}g \psi_{-}$ and $\psi^{\dag}_{-}g^{-1}\psi_{+}$.
Differentiating with respect to the sources and then setting them
equal to zero, we note that, as in eq.\ (\ref{2.25}), the correlators
factorise into a fermionic part which is a constant, and correlators
of $g$ in the WZW model.  Note that in the bosonic version of the
coset model as a gauge WZW model, correlators of tr($g$) are
constant \cite{gk}, which corresponds to correlators of
$\psi^{\dag}_{+} \psi_{-}$ in the above fermionic coset.

\section{Conclusions}

In summary we have given within the path integral framework a
complete derivation of the bosonization dictionary in 1+1
dimensions for both the abelian and non-abelian cases.  This was
achieved by inserting an appropriate identity into the
generating functional for free fermions to introduce bosonic degrees
of freedom.  Then, after a chiral change of variables, the generating
functional can be factorized in terms of a $G/G$-coset model partition
function, which is dynamically trivial, and a bosonic model partition
function.

We have illustrated our method with an external source $A_{\mu}$.
However, it is also applicable when $A_{\mu}$ is a background field or
a dynamical field, that is, when we integrate over $A_{\mu}$.  This
makes it possible to apply our method to interacting models like
QCD$_2$ or the chiral Gross-Neveu model.

A further interesting aspect concerns the bosonization BRST symmetry
disovered in ref.\ \cite{dns}.  This symmetry emerges quite
naturally in our approach which, as we have shown, allows one to
extend smooth bosonization to non-abelian systems in a
straightforward manner.

\vspace{.3cm}

{\it Acknowledgments:} A.N.T. acknowledges support from the Foundation
for Research Development and thanks the physics department of La Plata
for their hospitality during a recent visit.

\vspace{.3cm}

{\it Note added:} After completing this manuscript we became aware of
refs.\ \cite{bq,ds2} where similar results are obtained.

 \end{document}